\documentclass[
    ,final            
  ]
  {aipproc}

\layoutstyle{8x11single}

\def\alt{\mathrel{\hbox{\rlap{\hbox{\lower4pt\hbox{$\sim$}}}\hbox{$<$}}}}
\def\agt{\mathrel{\hbox{\rlap{\hbox{\lower4pt\hbox{$\sim$}}}\hbox{$>$}}}}
\def\lesssim{\mathrel{\hbox{\rlap{\hbox{\lower4pt\hbox{$\sim$}}}\hbox{$<$}}}}
\def\gtrsim{\mathrel{\hbox{\rlap{\hbox{\lower4pt\hbox{$\sim$}}}\hbox{$>$}}}}

\def\farcm{\hbox{$.\mkern-4mu^\prime$}}
\def\farcs{\hbox{$.\!\!^{\prime\prime}$}}

\def\ion#1#2{#1$\;${\small\rm\@Roman{#2}}\relax}

\def\simgt{\lower.5ex\hbox{$\; \buildrel > \over \sim \;$}}
\def\simlt{\lower.5ex\hbox{$\; \buildrel < \over \sim \;$}}

\def\etal{{et~al.}}
\def\amin{\ifmmode^{\prime}\else$^{\prime}$\fi}
\def\asec{\ifmmode^{\prime\prime}\else$^{\prime\prime}$\fi}

\def\simgt{\lower.5ex\hbox{$\; \buildrel > \over \sim \;$}}
\def\simlt{\lower.5ex\hbox{$\; \buildrel < \over \sim \;$}}

\newcommand\asca{{\it ASCA\/}}

\newcommand\chandra{{\it Chandra}}

\newcommand\xmm{{\it XMM\/}-Newton}

\newcommand\integral{{\it INTEGRAL}}

\newcommand\hess{{\it HESS}}
\newcommand\HESS{{\it HESS}}
\newcommand\magic{{\it MAGIC}}

\newcommand\glast{{\it GLAST}}

\newcommand\snr{G12.82${-}$0.02}
\newcommand\tev{\hbox{HESS~J1813$-$178}}
\newcommand\psr{CXOU J181335.17$-$174957.4}

\renewcommand\AIPsectionpreskip {10pt}
\renewcommand\AIPsectionpostskip {8pt}
\renewcommand\AIPsectionfont {\normalfont\fontsize{10}{12}\bfseries}

\begin{document}

\title{Discovery of a Pulsar Candidate Associated with the TeV Gamma-ray Source \tev}


\author{E.~V.~Gotthelf}{
  address={Columbia Astrophysics Laboratory, Columbia University, 550 West120$^{th}$ Street, New York, NY 10027, USA}
}

\author{D.~J.~Helfand}{
  address={Columbia Astrophysics Laboratory, Columbia University, 550 West120$^{th}$ Street, New York, NY 10027, USA}
}

\begin{abstract}

We present a \chandra\ X-ray observation of \snr, a shell-like radio
supernova remnant coincident with the TeV gamma-ray source \tev. We
resolve the X-ray emission from the co-located \asca\ source into a
compact object surrounded by structured diffuse emission that fills
the interior of the radio shell. The morphology of the diffuse
emission strongly resembles that of a pulsar wind nebula. The spectrum
of the compact source is well-characterized by a power-law with index
$\Gamma \approx 1.3$, typical of young and energetic rotation-powered
pulsars.  For a distance of 4.5 kpc, consistent with the X-ray
absorption, the 2--10~keV X-ray luminosity of the putative pulsar and
nebula is $L_{PSR} = 3.2 \times 10^{33}$~erg~s$^{-1}$ and $L_{PWN} =
1.4 \times 10^{34}$~erg~s$^{-1}$, respectively. Both the flux ratio of
$L_{PWN}/L_{PSR} = 4.3$ and the total luminosity of this system imply
a pulsar spin-down power of $\dot E > 10^{37}$~erg~s$^{-1}$, on a par
with the top ten most energetic young pulsars in the Galaxy. We
associate the putative pulsar with the radio remnant and the TeV source
and discuss the origin of the $\gamma$-ray emission.

\end{abstract}

\maketitle


\section{INTRODUCTION}

The \HESS\ observatory has opened an important new window onto the
highest energy astrophysical processes \hbox{($>10^{11}$~eV~$ =
0.1$~TeV)} with unprecedented sensitivity and spatial resolution
\cite{aha05}.  Of the 22 Galactic TeV sources detected by \hess\ over
the first two years of four-telescope operation, nearly half are
associated with supernova products -- supernova remnants (SNRs) or
pulsar wind nebulae (PWNe) \cite{aha06,fun06a}. With the exception of
two associations with binary systems, the remainder have yet to be
identified with a known source at any other wavelength.  The growing
number of supernova products associated \hess\ sources provide a
unique opportunity to locate new examples of these objects and to
determine the mechanism(s) for generating their $\gamma$-ray photons.

In this paper we focus on \snr, the first case of a SNR located by its
TeV emission \cite{bro05,hel05}.  This previously uncatalogued faint
shell-type radio supernova remnant (diameter $\sim 2^{\prime}$) lies
within the error circle of the unidentified source \tev\
\cite{aha05,aha06,fun07}. It was soon found to be coincident with an
unpublished non-thermal 2--10~keV \asca\ X-ray source
\cite{bro05,hel05}, a 10--100~keV \integral\ hard X-ray source
\cite{ube05}, and a 0.4--10~TeV \magic\ $\gamma$-ray detection
\cite{alb05}. Based on a recent follow-up \xmm\ X-ray observation, the
2--10~keV X-ray morphology suggests a composite remnant interacting
with ambient material to produce the TeV emission \cite{fun06b}.

Herein we report on a high-resolution X-ray imaging observation of
\snr\ obtained with the \chandra\ Observatory. This new data allows us
to associate this remnant with a young, energetic rotation-powered
pulsar generating a bright wind nebula, ultimately responsible for the
TeV emission from \tev.  In the following we use a distance to \snr\
of $d=4.5$~kpc as suggested by \citet{hel07}.

%
\begin{figure*}[t]
\small
\centerline{
\hfill
\includegraphics[width=1.95in,angle=270]{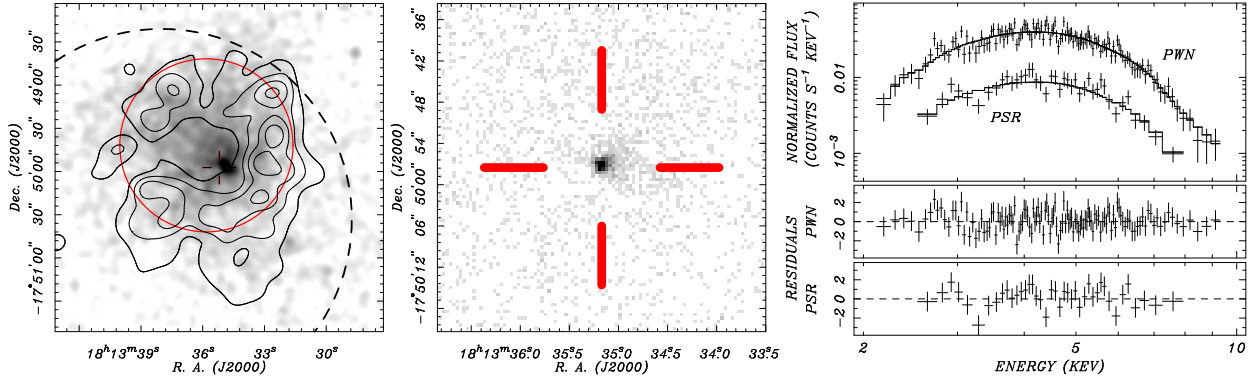}
\hfill
\includegraphics[width=1.95in,angle=270]{g12_acis_blowup.ps}
\hfill
\includegraphics[width=1.95in,angle=270]{g12_psr_pwn_spec.ps}
\hfill
}
\caption{\small A 30~ks \chandra\ ACIS observation of SNR \snr.  {\it
Left --} The broad-band smoothed X-ray image of \snr\ with the VLA radio
contours overlaid; the point source, whose location is indicated by
the cross, is removed to highlight the diffuse emission. The solid
circle ($d = 2^{\prime}$) illustrates that the radio shell encloses
the bulk of the X-ray emission. The dashed circle gives the $1\sigma$
extent of \tev. {\it Middle --} Blow up of the X-ray image centered on
the candidate pulsar, in unbinned CCD pixels.  The intensity is scaled
to highlight the point source and inner filamentary feature.  {The
cross is the same size (in arcsec) and location in both plots.}  {\it
Right --} ACIS spectra of both the putative pulsar (PSR) and pulsar
wind nebula (PWN). The solid lines show the best fit models given in
the text.  Residuals from these fits are shown in the lower panels. }
\label{fig1}
\end{figure*}

\section{THE CHANDRA X-RAY OBSERVATION OF SNR \snr}

A 30~ks X-ray observation of SNR \snr\ was obtained on 2006 Sept 15 UT
using the \chandra\ X-ray observatory (see \citet{hel07} for a full
description of this observation). Data were collected with the
Advanced CCD Imaging Spectrometer (ACIS) in the focal plane, operating
in the nominal full-frame TIMED/VFAINT exposure mode. The coordinates
of the \asca\ source associated with \snr\ were positioned on the
ACIS-I3 CCD chip and offset by $2^{\prime}$ from the nominal aimpoint
to allow any extended X-ray emission from the source to fall wholly on
this chip.  A total of 29.6~ks of live-time was accumulated with a CCD
frame time of 3.241~s (1.3\% deadtime).  We used the pipeline level 2
event products and reduced and analyzed this data using the standard
X-ray analysis software.

As shown in Figure~1, the \chandra\ ACIS image of \snr, with a near
on-axis spatial resolution of $\approx 0\farcs5$, fully resolves the
\asca\ source into diffuse X-ray emission and a point source.  The
diffuse flux generally fills in the radio shell and peaks toward the
point source emission, located at RA = 18:13:35.17, Dec = -17:49:57.48
(J2000; $1\sigma$ error radius of $\approx 0\farcs2$). These
coordinates were determined after registering the image using nine
X-ray field stars with USNO-B1 optical counterparts, and lies less
than $1^{\prime}$ from the maximum probability centroid of \tev, well
within its $2\farcm2$ ($1\sigma$) extent.

The point source is clearly offset from the center of the nearly
symmetric radio remnant, by $20^{\prime\prime}$.  This suggests either
rapid source motion or asymmetric SNR expansion relative to a common
birth center.  Extending west/southwest of the source is a faint
localized loop of emission not clearly resolved (Figure~1, middle
plot). A radial profile centered on the point source shows that this
excess emission is prominent out to $5^{\prime\prime}$ and disappears
into the background at about $10^{\prime\prime}$. The profile of the
point emission is consistent with the mirror response function when
the large-scale diffuse flux is taken into account. No obvious
evidence is found for any X-ray structure corresponding to the radio
shell.

To further study the nature of the X-ray emission from \snr\ we
extracted spectra from the point source, the diffuse nebula, and the
faint inner nebula, and from appropriate background regions.  A total
of 864 photons were collected from a $2\farcs0$ radius aperture
centered on \psr; this included $19$ background counts, estimated from
the local diffuse emission. Because of the low count rates involved,
photon pile-up is negligible. To encompass the bulk of the nebula,
we define an $80^{\prime\prime}$-radius region, centered on the
remnant, with the point source and inner nebula regions excluded (see
below); a total of 6638 photons were obtained. The background for this
spectrum was extracted from a $56^{\prime\prime}$-radius circle offset
due north of the nebula region and contained 1363 counts. For the
inner nebula we use a $6^{\prime\prime} \times\ 8^{\prime\prime}$
elliptical extraction region tilted with a position angle of
$240^\circ$ with the point source excluded. After accounting for the
local background (28\%) this resulted in a total of 223 photons.

Spectra from each region were grouped for a minimum of 15 cts per
spectral channel and fitted using the XSPEC software in the 2--10~keV
energy band, below which the line-of-sight flux is evidently highly
absorbed.  These CCD spectra ($\Delta E/E \sim 0.06$ FWHM at 1 keV)
are all well-characterized (but not uniquely) by an absorbed power-law
model associated with non-thermal emission. For the nebula flux, the
best-fit photon index is $\Gamma = 1.3(1.1,1.6)$, averaged over this
region, with an $N_H = 9.8(8.7-11) \times 10^{22}$~cm$^{-2}$,
consistent with that derived from NANTEN $^{12}CO(J$=1--0)
measurements \cite{fun07}.  In fitting the putative pulsar the column
density was fixed to the value derived for the higher-significance
nebula spectrum. The best-fit model yields a surprisingly similar
index, $\Gamma = 1.3^{1.6}_{1.0}$, not unlike that found for the Vela
pulsar, and to be expected for the most energetic rotation-powered
pulsars (cf. \citet{got03}). The absorbed 2--10~keV fluxes for the
putative pulsar (PSR) and PWN are $F_{PSR} = 1.3 \times
10^{-12}$~erg~cm$^{-2}$~s$^{-1}$ and $F_{PWN} = 5.6 \times
10^{-12}$~erg~cm$^{-2}$~s$^{-1}$, respectively. These results are
consistent with the \asca\ measurements of the composite spectrum
and, as discussed below, have important implications for the
energetics of the system.  The spectral distribution of the inner
nebula (IN) photons is not well-constrained, but a fit with the
power-law model yields a somewhat flatter index of $\Gamma =
0.4^{0.8}_{-0.3}$ than that of the PWN, with a flux of $F_{IN} \sim 4
\times 10^{-13}$~erg~cm$^{-2}$~s$^{-1}$.  Significantly, no emission
line features are found in any of the spectra analyzed.

\section{DISCUSSION}

The \chandra\ results offer a compelling case that the high-energy
emission from \snr\ is derived from the spin-down energy of a young
rotation-powered pulsar.  Deep radio pulsar searches, however, have so
far failed to detect a signal from \psr\ to an upper-limit that
approaches those measured for the least luminous young pulsars. Nor have
searches for slow periodicity using the X-ray data produced evidence
of a signal, although these results are not very constraining, with a
$3\sigma$ upper-limit for a sinusoidal modulation of $44\%$ for
\asca\ ($P>125$~ms), 27\% for ACIS ($P>6.5$~s), and 100\% for \xmm\
($P>147$~ms) \cite{bro05,hel07}.  More significant timing searches are
needed to verify our pulsar hypothesis for \snr. A detection will
allow a search for pulsed $\gamma$-rays with the \glast\
mission. Together, these results will help fully quantify the
energetics of this system.


That the putative pulsar is highly energetic is manifest by the large
flux ratio  of $F_{PWN}/F_{PSR} = 4.3$ in the 2--10~keV energy band;
only rotation-powered pulsars with spin-down energy loss rates above
$\dot E \approx 4 \times 10^{36}$~erg~s$^{-1}$ have a ratio this large
or greater \cite{got03}. Furthermore, the total 2--10~keV luminosity
of $L_X = 1.74 \times 10^{34} \ d^2_{4.5}$ erg s$^{-1}$ from \snr\ 
corresponds to $\dot E \sim 10^{37}$~erg s$^{-1}$ \cite{pos02},
placing this object among the Galaxy's top ten most energetic
pulsars. In particular, \snr\ bears a striking similarity in many
respects to SNR~G106.6+2.9, a radio remnant undetected in X-rays that
contains the energetic pulsar PSR~J2229+6114 with $\dot E = 1.8 \times
10^{37}$~erg~s$^{-1}$ and flux ratio of $F_{PWN}/F_{PSR} = 9$
\cite{hal01}.

The nature of the inner nebula resolved by the \chandra\ data is not
yet clear. If we assume that the pulsar's velocity vector is away from
the SNR center, it is hard to reconcile this feature with a
bow-shock origin. As a torus, it is unusually asymmetric and not well
centered on the pulsar. The relation between this feature and that of
the greater nebula requires a deeper \chandra\ observation to 
consider further.

The TeV emission from \tev\ is also likely to be also powered,
ultimately, by the putative pulsar. The spatially coincident X-ray and
TeV emission is notably rare among the \hess\ PWNs, likely indicating
a young ($<1000$~yr) pulsar, as in the Crab remnant. TeV emission for
older systems are considerably offset and may originate from ``relic''
electrons injected at an earlier epoch \cite{dja07}.  The
$\gamma$-rays from \snr\ could result from direct interaction of the
wind with the SNR shell, but no associated X-rays are seen.  Further
evidence for this is provided by the TeV/X-ray flux ratio of $L{\rm
(0.2-10 \ TeV)}/L{\rm (2-10 \ keV)} = 1.8$ -
although this ratio \cite{hel07} is the highest among the confirmed PWN
$\gamma$-ray emitters, it is well below that expected from the
interaction of an old SNR with a giant molecular cloud, as suggested
by Yamazaki \etal\ \cite{yam06}.  The TeV photons in this latter scenario
arise from $\pi^{\rm 0}$-decay of accelerated protons and the X-rays
are generated via synchrotron radiation from secondary electrons.

A more plausible explanation for the TeV emission from \snr\ is
inverse-Compton scattering of ambient photons off relativistic
electrons accelerated by the pulsar wind. Indeed, the remnant's
broadband spectrum (combining radio, X-ray, \integral, and \hess\
data) is well-fitted using the inverse-Compton model, although a large
photon field (1000 eV cm$^{-3}$) is required \cite{fun06b}. The nearby
HII region W33 may provides a significant local infrared photon flux,
but it is unlikely to reach the necessary photon density \cite{hel05}.
Given that several TeV detected PWNe lie near star-formation regions,
they may ultimately prove important for generating the TeV emission
\cite{hel07}.  Future \glast\ observation will be critical for
constraining possible spectral models and addressing these issues.


\bigskip\noindent 
This research is supported by grant GO6-7057X to EVG and by grant SAO~GO6-7052X to DJH.

\end{document}